# New Method for Public Key Distribution Based on Social Networks

Krzysztof Podlaski [1], Artur Hłobaż[1], Piotr Milczarski[1]

[1] Faculty of Physics and Applied Informatics, University of Lodz
`{podlaski,artur.hlobaz,piotr.milczarski}@uni.lodz.pl`

**Abstract.** The security of communication in everyday life becomes very important. On the other hand, all existing encryption protocols require from user additional knowledge end resources. In this paper we discuss the problem of public key distribution between interested parties. We propose to use a popular social media as a channel to publish public keys. This way of key distribution allows also easily connect owner of the key with real person institution (what is not always easy). Recognizing that the mobile devices become the main tool of communication, we present description of mobile application that uses proposed security methods.

## 1 Introduction

In nowadays people take into account security of information exchange. There are many different methods of encryption. The most spread and probably most popular are asymmetric methods based on a pair of users keys, public and private. While private key has to be kept very secret the public should be freely distributed between all interested parties and here we arrive at the big gap in used protocols. All known methodologies are very interested in securing the keys and authorization. We can sign the public key via well-known institutions (VeriSign, Comodo SSL, GlobalSign, etc.) and prove that a defined person or a company created this key. Even having a given key of a John Smith from Milwaukee how can a person be sure that this is exactly the same John Smith he knows? For many persons (even institutions) knowing their names and addresses is not enough. On the other hand if the John Smith is somebody's friend in real life he can be a "friend" in virtual one. They usually are connected via social network (facebook, linkedIn, etc.). The life would be much simpler if we could obtain his public key from this social network. In this paper we introduce an architecture for applications that allows sending encrypted information between two mobile devices using public key infrastructure and social media with QRcodes as a method of seamless distribution of public key.

The paper is organized as follows. In Section 2 we analyze the possibility of the storage of the public key with use of social media. The next Section contains requirements for QRcodes. Section 4 focuses on the used encryption method. Section 5 contains description of the proposed application architecture. At the end we present our final remarks and conclusions.

## 2 Public key distribution using social media

There are many interesting methods of public key distribution. One of the well-known methods of public key distribution is usage of key servers. Conventional PKI and PGP are still hard to be used by average users [1, 2]. The task to acquire valid public key of a friend is not an easy one. Nowadays users are used to use the social networks as the environment for searching any personal information. The everyday social networks and mobile devices revolutionized ways of communication. The average user is used to integrate all mobile devices with some social medias and requires all important data to be synchronized with the device phone book. Unfortunately existing key servers are not ready to be used in such a way. Some important elements have to be taken into account during the process of public key distribution:

a) ownership of the stored public key,
b) correspondence between the owner and real party (person, company, foundation, etc.),
c) easy accessibility to all interested parties,
d) is the key still actual/valid or was revoked.

Even though we have the key from some public storage in order use it we have to be sure to whom it belongs. The name of the person or company and even address are not always enough. Analyzing presented requirements we can notice that usual PKI or PGP key distribution does not always fulfill point b and d. We can try to use the webpage of a party or company, but there are often some additional problems:

- what the page address is,
- where the key is stored,
- how to obtain the key automatically.

On the other hand, nowadays the social media are the most spread and used mean of information distribution. Based on that experience there is an idea of using that medium for key distribution. First we have to analyze what kinds of information are already used in social medias. We can easily stress that on most of social portals users can store some data. The security measures used in such medias restrict that only the owner of the account can store and change this information. We can identify two types of information:

1. persistent data - like photos (usually more than one), web page address, email address,
2. transient data - like status, notes and memos.

The first type of information is usually stored in the users profile while the latter in some blog type medium. It is obvious that transient data is not a good candidate for our purpose. This means we should concentrate on elements that can be stored in users' profile. Moreover, it has to be noted that usually user is not allowed to customize what kind of information can be stored there.

It was already proposed by [3] to store link to our public key as one of user's web addresses. This is interesting idea however the user still has to have some special place for storing the key and social medium is used only as the information where to find the public key. Moreover, this method is easy accessible by machines while strange web addresses are not well perceived by humans.

The second very interesting place for the key is user profile photo/image or gallery (if exists). This will give us huge area for implementations if we would be able to store the public key inside the photo gallery. Now, we will try to cover this case more carefully. We have to take into account that social media usually optimize images that often means resizing, increasing jpg compressions.

### 2.1    Storing public key in photo metadata

Almost all image formats allow storing some additional information in attached metadata (Exif [4], XMP [5], IPTC [6]). That would be a good place to store the public key inside the metadata of profile picture. Unfortunately, the most known social portals (facebook, linkedIn) erase all metadata after the upload. This means that if we upload a photo with some information added in its metadata the information would be lost and not available for others.

### 2.2    Storing public key in photo file

There are many methods of steganography [7-8] that allow storing some information inside images. Unfortunately, these methods are very sensitive on operations like resizing and jpg optimization. All images uploaded to social medias are optimized and this operation would make impossible usage of steganography. Even hiding public key after the file closing marker would not work because all the information after the EOF (End Of File) marker is deleted by social media portal.

### 2.3    Storing public key inside as a QRcode

QRcode [9-10] is an image that encodes some text. It is possible to store public key as QRcode. The idea of using QRcode as key exchange for secure mobile communication was presented in our previous paper [11]. That way of storing the key has some advantages:

1. QRcode does not lose information during usual image resizing,
2. QRcode does not lose information during changes of format {.jpg, .png, .gif …}.

Storing public key as QRcode inside user gallery agrees with all requirements. We should decide for some nomenclature of naming the file with public key QRcode. Unfortunately, we cannot store it as profile picture, most of the people prefer to use real photo in that place. On the other hand, QRcodes nowadays are so widespread that they should not be perceived as out of place in user's photo gallery.

### 2.4 Storing QRcode on a given picture

There are some possibilities to store QRcode and a given picture together. There are methods like colored QRcodes but they are not acceptable for profile picture. There are also approaches to include a QRcode inside a picture. This is however not possible for all images and small QRcodes can be lost during resizing and optimization procedures. It is possible that some encoding of QRcode in image using HSB color space would be resistant for resizing and optimization but the impact of such procedure on image itself has to be determined.

### 2.5 Conclusions

According to presented analyze the best choice is to store public key in form of QRcode inside user's profile on social portal. This solves easily the problem of propagation and accessibility, on the other hand keeping some information about authenticity – only user can store photos in his/her gallery and prevents phishing attacks [12-13]. The problem of revoking the old key is easy to organize in proposed manner also. Moreover if somebody would like to narrow group of users that can view/use public key then access to galleries can be restricted to selected group of users (friends), this is possible in most of social portals.

## 3 Selection of QR code parameters

On the basis of [9-11] it was found that the best type of QR code to use for our purposes will be the version 17 (85x85). It will allow hiding a public key with the length of 2048 bits with the highest possible error correction feature – level H, approx. 30%. In the Table 1 shown below there is short description of QRcodes capacity using different variants of QRcodes.

**Table 1.** QR code types and their capacity

| Parameters | QR code type | | | |
|---|---|---|---|---|
| | QR code Model 1 | QR code Model 2 | Micro QR code | iQR code |
| Max Size [modules] | 73x73 | 177x177 | 17x17 | 422x422 |
| Max Capacity in numerals | 1101 | 7089 | 35 | 40637 |

| Max Capacity in alphanumeric | 667 | 4296 | 21 | ~24626 |
|---|---|---|---|---|
| Max Capacity in binary [bytes] | 458 | 2953 | 15 | ~16928 |

## 4 Encryption variants

Depending on the amount of data to be transferred between users, we can distinguish two possible encryption schemes [14, 15] which have application in mobile implementation described in the next section:

A. asymmetric cryptography,
B. asymmetric cryptography together with symmetric.

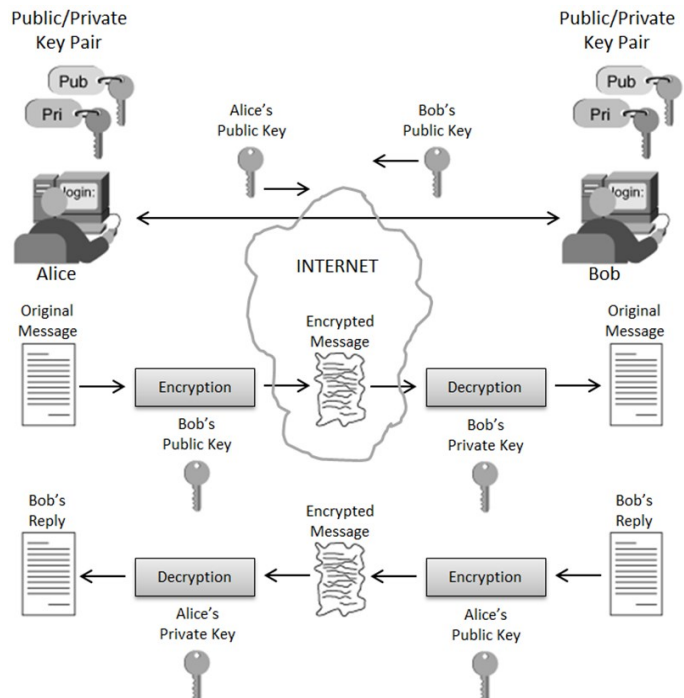

**Fig.1.** Asymmetric cryptography - ensuring data integrity and confidentiality

Because the asymmetric cryptography is slower than the symmetric one, first variant should only be applied to transmit short information, such as chat or SMS. If the user wants to send information, he encrypts it by the receiver's public key, which he has collected earlier from the social networking site. The receiver decrypts the message with its private key known only to him. Similarly, this is done the other way (Fig. 1).

One of the problems is that this scheme above does not provide authentication/identification of the information about the sending user. To ensure the authentication of the sender, the sender should first encrypt the message with its private key, and after that encrypt it again by the public key of the receiver. This allows the receiver be sure who is sending a message to him, because he will have to download the sender's public key from the social networking site in order to decrypt the information (Fig. 2).

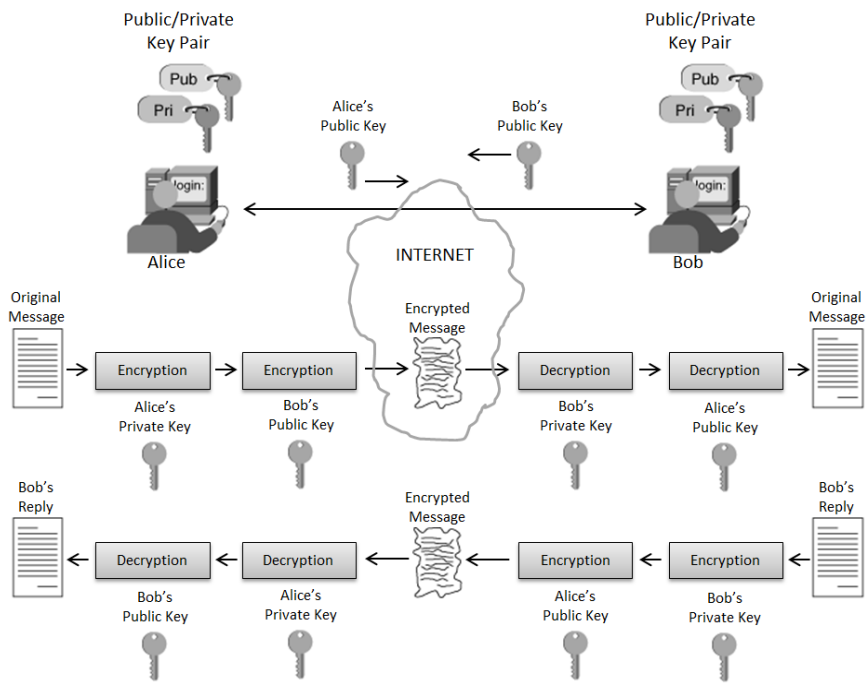

**Fig. 2.** Asymmetric cryptography - ensuring data integrity, confidentiality and sender authentication

In the case of the second encryption variant, the use of asymmetric cryptography with symmetric would allow to encrypt long information, i.e. files or stream call. In this variant, the transmission will be encrypted using symmetric cryptography. Exchange of components, which are important to establish a common one-time session key, will be done using asymmetric cryptography [16]. To establish a one-time session key, each party must first randomly generate 128 or 256-bit secret key. Secret key length depends on the used session encryption algorithm (AES-128 or AES-256). Then, the keys must be exchanged between parties using asymmetric cryptography (analogous manner as shown in Figure 1). At this point, each party has two secret keys. To establish a common one time session key each of the sides uses XOR operation on these two secret keys (Fig. 3).

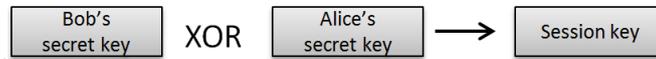
**Fig. 3.** The process of session key establishing

The secure information exchange process is shown in Fig. 4.

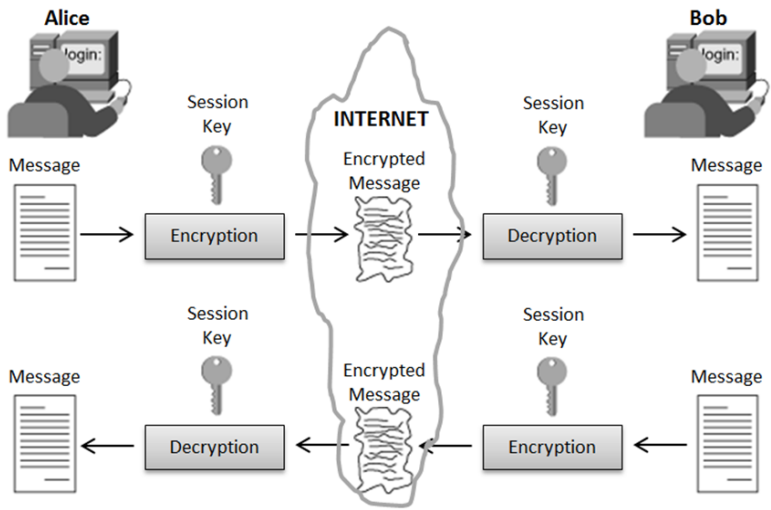

**Fig. 4.** Secure information exchange process

The proposed solution of session key establishment is different from standard methods like Diffie-Hellman algorithm. This means that no clear text information related to the key will not be possible to eavesdrop [17, 18]. The attacker will be able to capture only the data already encrypted with which he will not be able to do anything.

We should analyze the impact on the presented method when somebody breaks into user's social portal account (assume it is Bob's account). Even though such possibility exists the evil party could only change the user's public key into fake one. Such action would make impossible to continue encrypted communication between Alice and Bob because Bob's original private and fake public keys would not be paired anymore. Moreover, the intruder would not be able to decrypt messages sent by Alice until the Bob's private key is compromised. In the result Bob would be informed that his social account was broken. He would probably upload original public key once again and increase security measures of his social account. This implies that the main security precautions have to be taken when implementing proposed method. This will be important to keep private key secure in mobile device application.

## 5 Mobile implementation

### 5.1 Mobile communication

Usually encrypted communication begins with keys exchange. In proposed method of asymmetric cryptography suites perfectly for short messages and does not need any key exchange protocols at all. Interested parties can obtain appropriate receiver key from social network and stores own private one. Therefore, it can be easily deployed for smartphones to encrypt SMS communication. The latter of proposed methods need an exchange of a session encryption key and can be used on smartphones, tablets or even laptops (devices which have screen and camera) for stream communications.

### 5.2 Mobile application requirements

In order to implement mobile application that uses the presented encryption methods with usage of social network as public key store we define the application prerequisites:
- the Internet access,
- Social Medias Network access, can upload and download files/images from them,
- save data (keys),
- can generate Qrcode,
- process Qrcodes,
- can capture and send SMS, or capture voice telephony agent to work with the stream voice transmission,
- can encrypt and decrypt using presented methods.

The prerequisites of the application can be widened or shortened due to the application's functionality. The idea of application that uses text SMS can be shown on simple diagram (Fig. 5). The idea of the method and its mobile applications is presented in the paper [19].

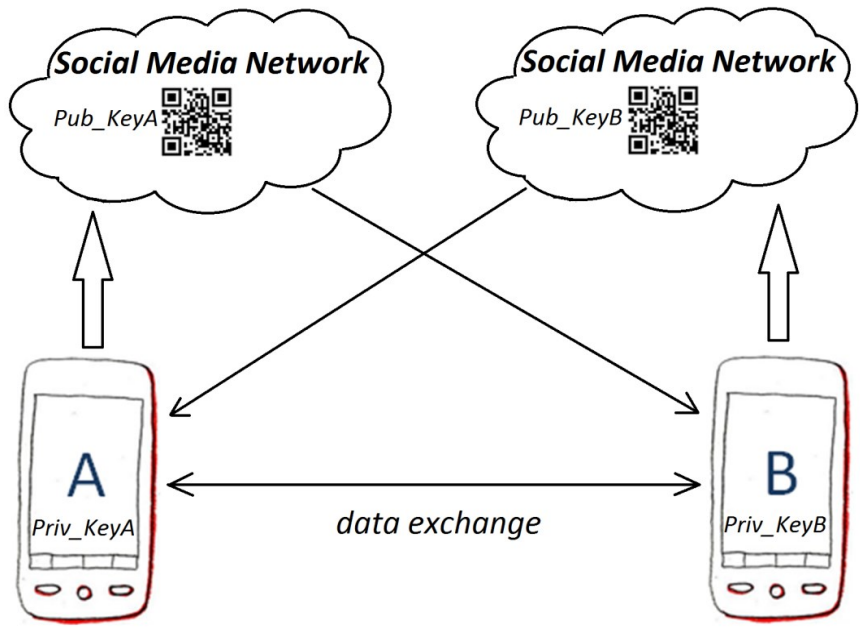

**Fig. 5.** SMS encrypting application that uses proposed cryptographic method

## 6      Conclusions

In nowadays there is need for secure data exchange and the mobile devices are the most spread tools used for communication. On the other hand most of users does not have enough skills to use sophisticated encryption methods and key exchange protocols. The need for secure communication and the ease of usage are very welcome by the community. The proposal of secure data exchange with everyday social network as key store solves both of the problems. On the other hand usage of images from social network gallery makes it easy to accept by ordinary mobile user. The proposed method can be implemented on mobile smartphones and description of such application was presented. The application will be presented in more detailed manner in next articles.